\pdfoutput=1

\documentclass[aps,prl,10pt,twocolumn,showpacs,superscriptaddress]{revtex4-1}
\usepackage{amsmath}
\usepackage{latexsym}
\usepackage{amssymb}
\usepackage{bm}
\usepackage{graphics,epstopdf}
\usepackage{color}
\usepackage{amsmath}

\usepackage[usenames,dvipsnames,svgnames]{pstricks}
\usepackage{epsfig}
\usepackage{pst-grad} 
\usepackage{pst-plot} 
\usepackage{epstopdf}
\usepackage{newlfont}
\usepackage{amssymb}
\usepackage{amsfonts}
\usepackage{amsmath}
\usepackage{amsthm}
\usepackage{graphicx}
\usepackage{epsfig}
\usepackage{color}
\usepackage{bm}
\usepackage[breaklinks]{hyperref}
\usepackage{tikz}

\usepackage{times}

\begin{document}
\title{Tighter
Uncertainty and Reverse Uncertainty Relations}
\affiliation{Quantum Information and Computation Group,\\
Harish-Chandra Research Institute, Chhatnag Road, Jhunsi, Allahabad, India\\and}
\affiliation{Homi Bhabha National Institute,
Anushaktinagar, \\Training School Complex, Mumbai-400085, India}
\author{Debasis Mondal}
\email{debamondal@hri.res.in}
\author{Shrobona Bagchi}
\email{shrobona@hri.res.in}
\author{Arun Kumar Pati}
\email{akpati@hri.res.in}


\date{\today}

\begin{abstract}
 We prove a few novel state-dependent uncertainty relations for product as well the sum of variances of two incompatible observables. These uncertainty relations are shown to be tighter than the Roberson-Schr{\"o}dinger uncertainty relation and other ones existing in the current literature. Also, we derive state dependent upper bound to the sum and the product of variances using the reverse Cauchy-Schwarz inequality and the Dunkl-Williams inequality. Our results suggest that not only we cannot prepare quantum states for which two incompatible observables can have sharp values, but also we have both, lower and upper limits on the variances of quantum mechanical observables at a fundamental level.
 
\end{abstract}

\maketitle

{\em Introduction.---}Quantum mechanics has many distinguishing features from classical mechanics in the microscopic world. One of these distinguished features is the existence of incompatible observables. 
As a result of this incompatibility, we have the uncertainty principle and the uncertainty relations. Owing to the seminal works by Heisenberg \cite{heisenber}, Robertson \cite{robert} and Schr{\"o}dinger \cite{Schrodinger}, lower bounds were shown to exist for the product of variances of two non-commuting observables. Recently, Maccone and Pati have shown stronger uncertainty relations for all incompatible observables \cite{maccone}. The stronger uncertainty relations have also been experimentally tested \cite{kunwang}.
In addition, the entropic uncertainty and the reverse uncertainty relations also capture the essence of quantum uncertainty \cite{maassen,deutsch,huang,jorge,jorge1,puchala,jorge,jorge1}  and the incompatibility between two observables, but in a state-independently way.

 With the advent of quantum information theory, uncertainty relations in particular, have been established as important tools for a wide range of
applications. To name a few, uncertainty relations have been used in formulating quantum mechanics \cite{busch} (where we can justify the complex structure of the Hilbert space \cite{lahti} or as a
fundamental building block for quantum mechanics and
quantum gravity \cite{hall}). Further, it has been used in entanglement detection \cite{guhne,hofmann}, security analysis of quantum key distribution in
quantum cryptography \cite{fuchs}, quantum metrology and quantum speed limit (QSL) \cite{deba,deba1,mandel,gerardospeed}. In most of these areas, particularly, in quantum entanglement detection and quantum metrology or quantum speed limit, where a small fluctuation in an unknown parameter of the state of the system is needed to detect, state-independent relations are not very useful. Thus, a focus on the study of the state dependent and tighter uncertainty and the reverse uncertainty relations based on the variance is a need of the hour.

 Uncertainty relations in terms of variances of incompatible observables are generally expressed in two forms--- product form and sum form. Although, both of these kinds of uncertainty relations express limitations in the possible preparations of the system by giving a lower bound to the product or sum of the variances of two observables, product form cannot capture the concept of incompatibility of observables properly because it may become trivial even when observables do not commute. In this sense, uncertainty relations in terms of the sum of variances capture the concept of incompatibility more accurately \cite{maccone}. It may be noted that earlier uncertainty
relations that provide a bound to the sum of the
variances comprise a lower bound in terms of the variance
of the sum of observables \cite{unpati,maccone}, entropic uncertainty and reverse uncertainty relations \cite{maassen,deutsch,huang,jorge,jorge1,puchala}, sum uncertainty
relation for angular momentum observables \cite{rivas}, sum uncertainty relations for N-incompatible observables \cite{fei2} and also uncertainty for non-Hermitian operators \cite{Bagchi, patiuttam,Hall1}. Recently, experiments have also been performed to test various uncertainty relations \cite{fei1,kunwang,baek}.

One striking feature of the most of the stronger uncertainty bounds is that they depend on arbitrary orthogonal state $|\Psi_\perp\rangle$ to the state of the system $|\Psi\rangle$ \cite{maccone,song,yao,xiaojing,chiruuncer}. It has been shown that an optimization of over $|\Psi_\perp\rangle$, which maximizes the lower bound, will saturate
the inequality. For higher dimensional systems, finding such an orthogonal state, may be difficult. Therefore, a focus on to derive an uncertainty relation independent of any optimization and yet tight is needed for the sake of further technological developments and explorations, particularly in quantum metrology \cite{seth}. Here, we aim to attain this goal and report a few tighter as well as optimization free uncertainty bounds both in the sum and the product forms.


The aim of this letter is two fold.
First, we show a set of uncertainty relations in product as well as sum forms. The new uncertainty relation in the product form is stronger than the Robertson-Schr{\" o}dinger uncertainty relation. We also derive an optimization free bound, which is also tighter most of the times than the Robertson-Schr{\"o}dinger relation. On the other hand, uncertainty relations for the sum of variances are also shown to be tight enough considering the advantage that the bounds do not need an optimization. Second, we prove reverse uncertainty relations for incompatible observables. We derive the state dependent reverse uncertainty relations in terms of variances both in the sum form and the product form. Thus, the uncertainty relation is not the only distinguishing feature but here, we show that reverse uncertainty relation also comes out as an another unique feature of quantum mechanics. If one considers that  uncertainty
relation quantitatively expresses the impossibility
of jointly sharp preparation of incompatible observables, then the reverse uncertainty relation should express the maximum extent to which the joint sharp preparation of incompatible observables is impossible. It is well known that quantum mechanics sets the lower limit to the time of quantum evolutions. In contrast to this, it is now expected that our state dependent reverse uncertainty relations may also be useful in setting an upper time limit of quantum evolutions \cite{uprep} (reverse bound to the QSL) and in quantum metrology. Thus, the results of our paper are not only of fundamental interest, but can have several applications in diverse areas of quantum physics, quantum information and quantum technology.

 
{\em Tighter uncertainty relations.---}
For any two non-commuting operators $A$ and $B$, the Robertson-Schr{\"o}dinger uncertainty relation \cite{Schrodinger} for the state of the system $|\Psi\rangle$ is given by the following inequality
\begin{eqnarray}\label{sun}
\hspace{-.2cm}\Delta A^2\Delta B^2\geq\left|\frac{1}{2}\langle[A,B]\rangle\right|
^2+\left|\frac{1}{2}\langle \{A,B\}\rangle-
\langle A\rangle\langle B\rangle\right|^2,
\end{eqnarray}
where the averages and the variances are defined over the state of the system $|\Psi\rangle$. This relation is a direct consequence of the Cauchy-Schwarz inequality. However, this uncertainty bound is not optimal. There have been several attempts to tighten the bound \cite{maccone,unpati,song,yao}. Here, we provide a tighter bound and obtain a new uncertainty relation. Let us consider two observables $A$ and $B$ in their eigenbasis as $A=\sum_{i}a_{i}|a_{i}\rangle\langle a_{i}|$ and $B=\sum_{i}b_{i}|b_{i}\rangle\langle b_{i}|$. Let us define $(A-\langle A\rangle)=\overline{A}=\sum_{i}\tilde{a}_i|a_i\rangle\langle a_i|$ and $(B-\langle B\rangle)=\overline{B}=\sum_{i}\tilde{b}_i|b_i\rangle\langle b_i|$. We express $|f\rangle=\overline{A}|\Psi\rangle$ and $|g\rangle=\overline{B}|\Psi\rangle$ as $|f\rangle=\sum_{n}\alpha_{n}|\psi_{n}\rangle$ and $|g\rangle=\sum_{n}\beta_{n}|\psi_{n}\rangle$, where $\{|\psi_{n}\rangle\}$ is an arbitrary complete basis. Using the Cauchy-Schwarz inequality  for two real vectors $\vec{\alpha}=\Big(|\alpha_{1}|,|\alpha_{2}|,|\alpha_{3}|,...\Big)$, $\vec{\beta}=\Big(|\beta_{1}|,|\beta_{2}|,|\beta_{3}|,...\Big)$, we have
\begin{eqnarray}\label{nform}
\Delta A^{2}\Delta B^{2}&=&\langle f|f\rangle\langle g|g\rangle=\sum_{n,m}|\alpha_{n}|^2|\beta_{m}|^2\nonumber \\&\geq & \Big(\sum_{n}|\alpha_{n}||\beta_{n}|\Big)^2=\Big(\sum_{n}|\alpha_{n}^{*}\beta_{n}|\Big)^2 \nonumber \\ &=&\Big(\sum_{n}|\langle\Psi|\overline{A}|\psi_{n}\rangle\langle\psi_{n}|\overline{B}|\Psi\rangle|\Big)^2\nonumber\\&=&\Big(\sum_{n}|\langle\Psi|\overline{A}~\overline{B}_{n}^{\psi}|\Psi\rangle|\Big)^2,
\end{eqnarray}
 where $\overline{B}_{n}^{\psi}=|\psi_{n}\rangle\langle\psi_{n}|\overline{B}$, $\alpha_{n}=\langle\psi_{n}|\overline{A}|\Psi\rangle$ and 
$\beta_{n}=\langle\psi_{n}|\overline{B}|\Psi\rangle$. On expressing $\langle\Psi|
\overline{A}~\overline{B}_{n}^{\psi}|\Psi\rangle=\frac{1}{2}\big{(}\langle[\overline{A},
\overline{B}_{n}^{\psi}]\rangle_{\Psi}+\langle\{\overline{A},
\overline{B}_{n}^{\psi}\}\rangle_{\Psi}\big{)}$, the new uncertainty relation can be written as
\begin{equation}\label{prodcomm}
\Delta A^{2}\Delta B^{2}\geq\frac{1}{4}\Big{(}\sum_{n}\Big|\langle[\overline{A},\overline{B}_{n}^{\psi}]\rangle_{\Psi}+\langle\{\overline{A},\overline{B}_{n}^{\psi}\}\rangle_{\Psi}\Big|\Big{)}^2.
\end{equation}
The new uncertainty relation is tighter than the Robertson-Schr{\"o}dinger uncertainty relation \cite{robert,Schrodinger} given in Eq. (\ref{sun}). To prove this let us start with the right hand side of Eq. (\ref{nform}) and note that
\begin{eqnarray}\label{prodcomp}
&\Big(&\sum_{n}|\langle\Psi|\overline{A}~\overline{B}_{n}^{\psi}|\Psi\rangle|\Big)^2\geq\Big|\sum_{n}\langle\Psi|\overline{A}~\overline{B}_{n}^{\psi}|\Psi\rangle\Big|^2\nonumber\\&=&\Big|\langle\Psi|\overline{A}~\overline{B}|\Psi\rangle\Big|^2,
\end{eqnarray}
where we have used the fact that $|\sum_{i}z_{i}|^2\leq (\sum_{i}|z_{i}|)^2$, $z_{i}\in\mathbb{C}$ for all i. Here, the last line in Eq. (\ref{prodcomp}) is nothing but the bound obtained in Eq. (\ref{sun}). Thus, our bound is indeed tighter than the Robertson-Schr{\"o}dinger uncertainty relation.

This uncertainty relation in Eq. (\ref{prodcomp}) can further be tightened by optimizing over the sets of complete orthonormal bases as 
\begin{equation}\label{prodcomm1}
\Delta A^{2}\Delta B^{2}\geq\max_{\{|\psi_{n}\rangle\}}\frac{1}{4}\Big{(}\sum_{n}\Big|\langle[\overline{A},\overline{B}_{n}^{\psi}]\rangle_{\Psi}+\langle\{\overline{A},\overline{B}_{n}^{\psi}\}\rangle_{\Psi}\Big|\Big{)}^2.
\end{equation}
\begin{figure}
\includegraphics[scale=0.85]{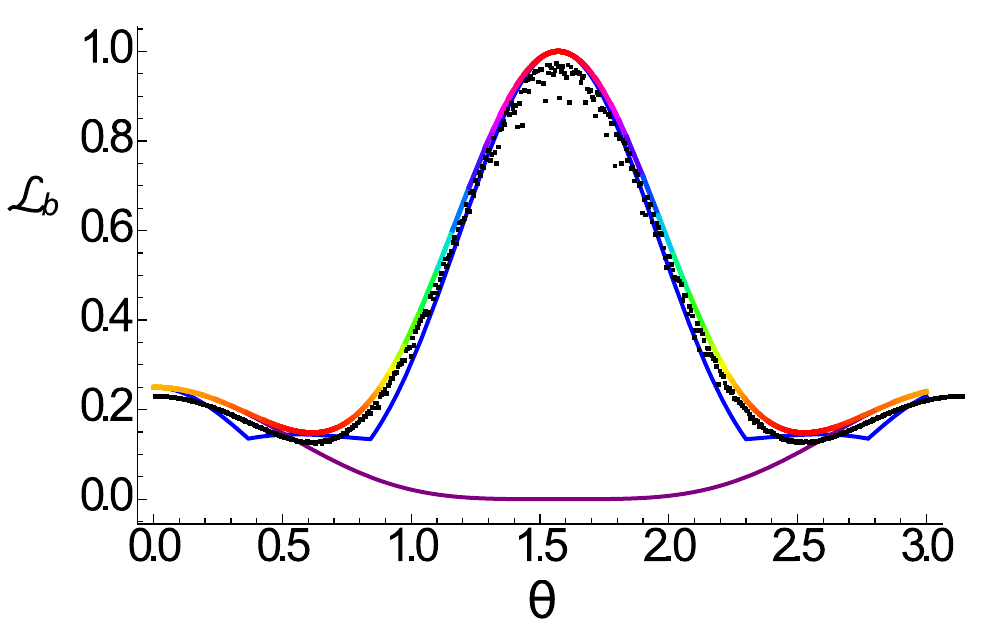}
\caption{\footnotesize Here, we plot the lower bound of the product of variances of two incompatible observables, $A=L_x$ and $B=L_y$, two components of the angular momentum for spin 1 particle with a state $|\Psi\rangle=\cos\theta|1\rangle-\sin\theta|0\rangle$, where the state $|1\rangle$ and $|0\rangle$ are the eigenstates of $L_{z}$ corresponding to eigenvalues 1 and 0 respectively. The blue line shows the lower bound of the product of variances given by (\ref{ww}), the purple coloured plot stands for the bound given by Schr{\"o}dinger uncertainty relation given by Eq. (\ref{sun}) and the hue plot denotes the product of two variances. Scattered black points denote the optimized uncertainty bound achieved by Eq. (\ref{prodcomm1}).}
\label{fig1}
\end{figure}
As shown in Fig. (\ref{fig1}), an optimization over different bases indeed gives tighter bound.

 Next, we derive an optimization-free uncertainty relation for two incompatible observables. For that we consider (say) $\overline{A}^2=\sum_{i,j}(a_i-a_jF_{\Psi}^{a_{j}})^2|a_i\rangle\langle a_i|=\sum_{i}(\tilde{a}_{i})^{2}|a_i\rangle\langle a_i|$ and $\overline{B}^{2}=\sum_{i,j}(b_i-b_jF_{\Psi}^{b_{j}})^2|b_i\rangle\langle b_i|=\sum_{i}(\tilde{b}_{i})^{2}|b_i\rangle\langle b_i|$, where $F_{\Psi}^{x}$ is nothing but the fidelity between the state $|\Psi\rangle$ and $|x\rangle$ $(|x\rangle=|a_i\rangle, |b_i\rangle)$, $F(|\Psi\rangle,|x\rangle)=|\langle\Psi|x\rangle|^2$. Using the Cauchy-Schwarz inequality, we obtain 
\begin{eqnarray}\label{ww}
&&\Delta A^2\Delta B^2\geq   
\Bigg(\sum_{i}\sqrt{F_{\Psi}^{a_{i}}}\sqrt{F_{\Psi}^{b_{i}}}
\tilde{a}_{i}\tilde{b}_{i}\Bigg)^2,
\end{eqnarray}
where we use the inequality for two real vectors $\vec{u}$ and $\vec{v}$ defined as $\vec{u}=\Big(\tilde{a}_{1}\sqrt{F^{a_{1}}_{\Psi}},\tilde{a}_{2}\sqrt{F^{a_{2}}_{\Psi}},\tilde{a}_{3}\sqrt{F^{a_{3}}_{\Psi}},...\Big)$, $\vec{v}=\Big(\tilde{b}_{1}\sqrt{F^{b_{1}}_{\Psi}},\tilde{b}_{2}\sqrt{F^{b_{2}}_{\Psi}},\tilde{b}_{3}\sqrt{F^{b_{3}}_{\Psi}},...\Big)$ and the quantities $\sqrt{F_{\Psi}^{a_{i}}}\tilde{a}_{i}$, $\sqrt{F_{\Psi}^{b_{i}}}\tilde{b}_{i}$ are arranged such that $\sqrt{F_{\Psi}^{a_{i+1}}}\tilde{a}_{i+1}\geq \sqrt{F_{\Psi}^{a_{i}}}\tilde{a}_{i}$ and $\sqrt{F_{\Psi}^{b_{i+1}}}\tilde{b}_{i+
1}\geq \sqrt{F_{\Psi}^{b_{i}}}\tilde{b}_{i}$. This new uncertainty relation depends on the transition probability between the state of the system and the eigenbases of the observables. The incompatibility is captured here not by the non-commutativity, rather by the non-orthogonality of the state of the system $|\Psi\rangle$ and the eigenbases of the observables $|a_{i}\rangle$ and $|b_{i}\rangle$.
\begin{figure}
\includegraphics[scale=0.85]{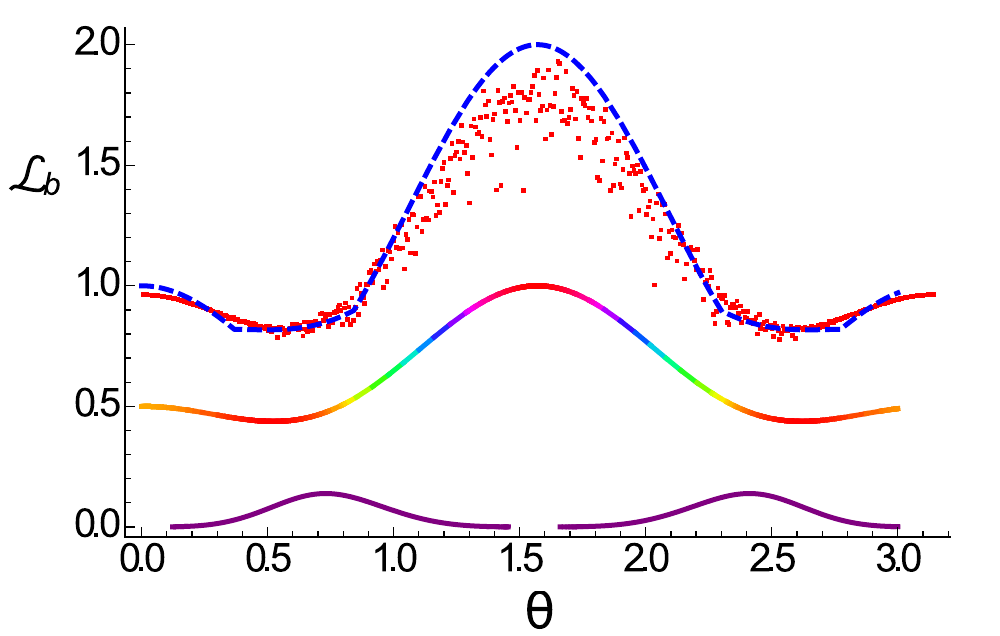}
\caption{\footnotesize Here, we plot the lower bound of the sum of variances for two incompatible observables, $A=L_x$ and $B=L_y$, two components of the angular momentum for spin-1 particle with a state $|\Psi\rangle=\cos\theta|1\rangle-\sin\theta|0\rangle$, where the state $|1\rangle$ and $|0\rangle$ are the eigenstates of $L_{z}$ corresponding to eigenvalues 1 and 0 respectively. Blue dashed line shows the lower bound of the sum of variances given by (\ref{parallelogram}), hue plot denotes the bound given by Eq. (4) in \cite{maccone} and the purple coloured plot stands for the bound given by Eq. (2) in \cite{unpati}. Scattered red points are the uncertainty bound achieved by Eq. (3) in \cite{maccone}. As observed from the plot, the bound given by Eq. (\ref{parallelogram}) is one of the tightest bounds in the literature. The bound given by Eq. (3) in \cite{maccone} is the only bound, which surpasses at only few points.}
\label{fig2}
\end{figure} 
  As observed from the Fig. (\ref{fig1}), the bound given by Eq. (\ref{prodcomm1}) is one of the tightest bounds reported here but it needs optimization. The bound given by Eq. (\ref{ww}) is the only bound, which is tighter than the other bounds most of the time and even surpasses the bound given by Eq. (\ref{prodcomm1}) yet it does not need any optimization. 
  
   However, we know that the product of variances does not fully capture the uncertainty for two incompatible observables, since if the state of the system is an eigenstate of one of the observables, then the product of the uncertainties vanishes. To overcome this shortcoming, the sum of variances was invoked to capture the uncertainty of two incompatible observables. In this regard, stronger uncertainty relations for all incompatible observables were proposed in Ref. \cite{maccone}. But, these uncertainty relations are not always tight and highly dependent on the states perpendicular to the chosen state of the system. Here, we propose new uncertainty relations that perform better than the existing bounds and need no optimization. We use the the parallelogram law for two real vectors to improve the bound on the sum of variances for two incompatible observables.
Using the parallelogram law for two real vectors $\vec{u}$ and $\vec{v}$, one can derive a lower bound on the sum of variances of two observables as
\begin{eqnarray}\label{parallelogram}
\Delta A^2+\Delta B^2
&\geq &\frac{1}{2}\sum_{i}\Big(\tilde{a}_{i}\sqrt{F_{\Psi}^{a_{i}}}+
\tilde{b}_i\sqrt{F_{\Psi}^{b_{i}}}\Big)^2.
\end{eqnarray}
As shown in Fig. (\ref{fig2}), the bound obtained in Eq. (\ref{parallelogram}) is one of the tightest optimization free bound.
 
 If one allows the optimization over a set of states, then the procedure used to derive the uncertainty relation given in Eq. (\ref{prodcomm1}) can be used to derive another set of uncertainty relations using the parallelogram law for two real vectors $\vec{\alpha}$ and $\vec{\beta}$.
Using the parallelogram law, one obtains
\begin{eqnarray}
\Delta A^2+\Delta B^2&\geq & \frac{1}{2}\sum_{n}\Big(|\alpha_{n}|+|\beta_{n}|\Big)^2\nonumber\\&=&\frac{1}{2}\sum_{n}\Big(|\langle\psi_{n}|\overline{A}|\Psi\rangle|+|\langle\psi_{n}|\overline{B}|\Psi\rangle|\Big)^2.
\end{eqnarray}
An optimization over the set of complete bases provides a more tighter bound as
\begin{equation}\label{{uoptpara}}
\Delta A^2+\Delta B^2\geq\max_{\{|\psi_{n}\rangle\}}\frac{1}{2}\sum_{n}\Big(|\langle\psi_{n}|\overline{A}|\Psi\rangle|+|\langle\psi_{n}|\overline{B}|\Psi\rangle|\Big)^2.
\end{equation}
{\em Reverse uncertainty relations.---} Does quantum mechanics restrict upper limit to the product and sum of variances of two incompatible observables? Here, for the first time, we introduce the reverse bound, i.e., the upper bound to the product and the sum of variances of two incompatible observables. To prove the reverse uncertainty relation for the product of variances of two observables, we use the reverse Cauchy-Schwarz inequality for positive real numbers \cite{rcs,rcs1,rcs2, Dunkl}. This states that \textit{for two sets of positive real numbers $c_{1},...,c_{n}$ and $d_1,...d_n$, if $0< c\leq c_{i}\leq C<\infty$, $0<d\leq d_{i}\leq D<\infty$ for some constants $c$, $d$, $C$ and $D$ for all $i=1,...n$, then }
\begin{figure}
\includegraphics[scale=0.85]{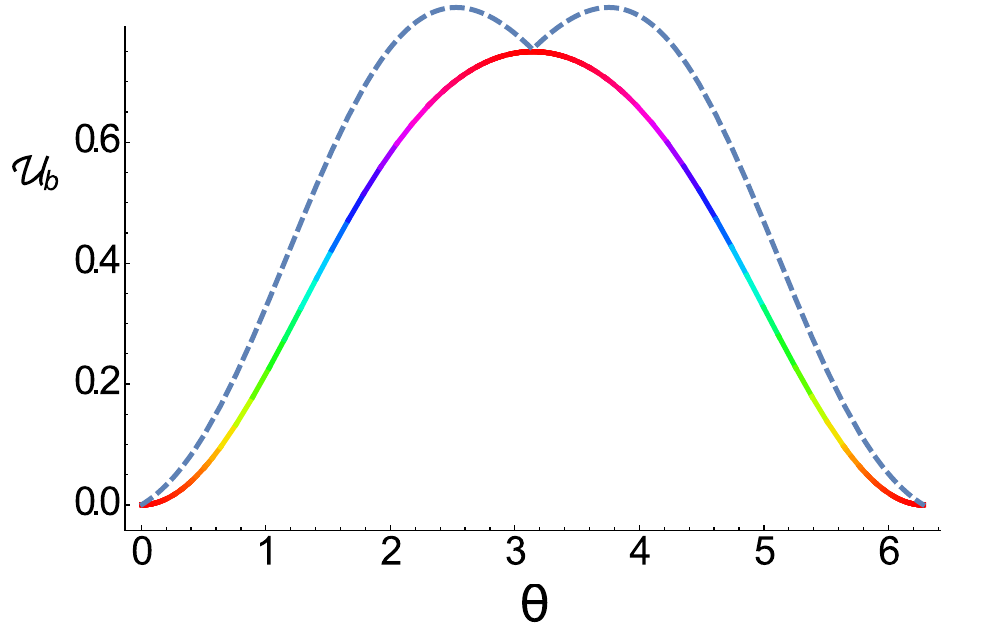}
\caption{\footnotesize Here, we plot the upper bound of the product of variances for two incompatible observables, $A=\sigma_x$ and $B=\sigma_z$, two components of the angular momentum for spin $\frac{1}{2}$ particle with a state $\rho=\frac{1}{2}\left(I_{2}+\cos\frac{\theta}{2}\sigma_{x}+\frac{\sqrt{3}}{2}\sin\frac{\theta}{2}\sigma_{2}+\frac{1}{2}\sin\frac{\theta}{2}\sigma_{z}\right)$. Blue dashed line is the upper bound of the product of the two variances given by (\ref{runp}) and the hue plot denotes the product of the two variances.}
\label{fig3}
\end{figure}
\begin{equation}
\sum_{i,j}c_{i}^2d_{j}^2\leq\frac{(CD+cd)^2}{4cdCD}\Big(\sum_{i}c_{i}d_{i}\Big)^2.
\end{equation}

Using this inequality for $c_{i}=\sqrt{F_{\Psi}^{a_{i}}}|\tilde{a}_{i}|$ and $d_{i}=\sqrt{F_{\Psi}^{b_{i}}}|\tilde{b}_{i}|$, one can show that the product of variances of two observables satisfies the relation
\begin{equation}\label{runp}
\Delta A^{2}\Delta B^{2}\leq\Omega_{ab}^{\Psi}\Bigg(\sum_{i}\sqrt{F_{\Psi}^{a_{i}}}\sqrt{F_{\Psi}^{b_{i}}}
|\tilde{a}_{i}||\tilde{b}_{i}|\Bigg)^2,
\end{equation}
where $\Omega_{ab}^{\Psi}=\frac{\big(M^{a}_{\Psi}M^{b}_{\Psi}+m^{a}_{\Psi}m^{b}_{\Psi}
\big)^2}{4M^{a}_{\Psi}M^{b}_{\Psi}m^{a}_{\Psi}m^{b}_{\Psi}}$ with $M_{\Psi}^{a}=\max\{\sqrt{F_{\Psi}^{a_{i}}}|\tilde{a}_{i}|\}$, $m_{\Psi}^{a}=\min\{\sqrt{F_{\Psi}^{a_{i}}}|\tilde{a}_{i}|\}$, $M_{\Psi}^{b}=\max\{\sqrt{F_{\Psi}^{b_{i}}}|\tilde{b}_{i}|\}$ and $m_{\Psi}^{b}=\min\{\sqrt{F_{\Psi}^{b_{i}}}|\tilde{b}_{i}|\}$.
If one uses the reverse Cauchy-Schwarz inequality for the two real positive vectors $\vec{\alpha}$ and $\vec{\beta}$, we have
\begin{eqnarray}\label{oprev}
&\Delta A^{2}&\Delta B^{2}\leq\Lambda_{\alpha\beta}^{\psi\Psi}\Bigg(\sum_{n}|\alpha_{n}||\beta_{n}|\Bigg)^2\nonumber\\&=&
\frac{\Lambda_{\alpha\beta}^{\psi\Psi}}{4}\Bigg(\sum_{n}\Big|\langle[~\overline{A},\overline{B}_{n}^{\psi}]\rangle+\langle\{\overline{A},\overline{B}_{n}^{\psi}\}\rangle\Big|\Bigg)^2,
\end{eqnarray}
where $\Lambda_{\alpha\beta}^{\psi\Psi}=\frac{\big(M^{\alpha}_{\psi\Psi}M^{\beta}
_{\psi\Psi}+m^{\alpha}_{\psi\Psi}m^{\beta}_{\psi\Psi}
\big)^2}{4M^{\alpha}_{\psi\Psi}M^{\beta}_{\psi\Psi}m^{\alpha}_{\psi\Psi}m^{\beta}
_{\psi\Psi}}$ with $M^{\alpha}_{\psi\Psi}=\max\{|\alpha_{n}|\}$, $m^{\alpha}_{\psi\Psi}=\min\{|\alpha_{n}|\}$, $M^{\beta}_{\psi\Psi}=\max\{|\beta_{n}|\}$ and $m^{\beta}_{\psi\Psi}=\min\{|\beta_{n}|\}$. One can optimize the right hand side of the Eq. (\ref{oprev}) to get a tighter reverse uncertainty relation.

Next, we derive the reverse uncertainty relation for the sum of variances using the Dunkl-Williams inequality \cite{Dunkl}. It is a state dependent upper bound on the sum of variances. The Dunkl-Williams inequality states that
\textit{if f, g are non-null vectors in the real or complex inner product space, then}
\begin{equation}
||f-g||\geq\frac{1}{2}(||f||+||g||)||\frac{f}{||f||}-\frac{g}{||g||}||.
\end{equation}
 Now, if we take $|f\rangle=\overline{A}|\Psi\rangle$ and $|g\rangle=\overline{B}|\Psi\rangle$ as defined earlier, then, using the Dunkl-Williams inequality we obtain the following equation 
 \begin{equation}\label{usumdev}
 \Delta A+\Delta B\leq \frac{\sqrt{2}\Delta (A-B)}{\sqrt{1-\frac{Cov(A,B)}{\Delta A.\Delta B}}},
 \end{equation} where, $Cov(A,B)=\frac{1}{2}\langle\{ A,B\}\rangle-\langle A\rangle\langle B\rangle$ is the \textit{quantum covariance} of the operators $A$ and $B$ in quantum state $|\Psi\rangle$.
 We know from the Robertson-Schr{\"o}dinger uncertainty relation that $\Delta A^2\Delta B^2\geq Cov(A,B)^2 +\frac{1}{4}|\langle [A,B]\rangle|^2$, such that $-1\leq\frac{Cov(A,B)}{\Delta A\Delta B}\leq 1$. Thus, the quantity inside the square root in the denominator of the Eq. (\ref{usumdev}) is always positive. Also, $ \sqrt{1-\frac{Cov(A,B)}{\Delta A.\Delta B}}< \sqrt{2}$ for non-trivial cases. Thus, we have $\Delta A+\Delta B< \Delta (A-B)$ in such cases, though this is a weaker bound than Eq. (\ref{usumdev}). Therefore,
by squaring the both sides of the equation, we obtain an upper bound on the sum of variances as 
\begin{figure}
\includegraphics[scale=0.85]{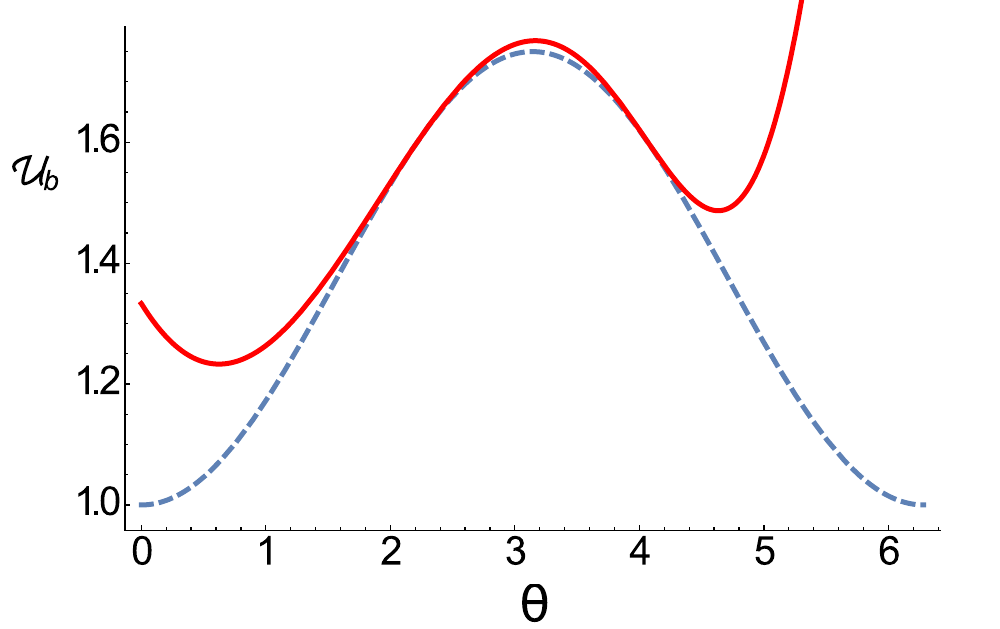}
\caption{\footnotesize Here, we plot the upper bound of the sum of variances for two incompatible observables, $A=\sigma_x$ and $B=\sigma_z$, two components of the spin angular momentum for spin $\frac{1}{2}$ particle with a state $\rho=\frac{1}{2}\left(I_{2}+\cos\frac{\theta}{2}\sigma_{x}+\frac{\sqrt{3}}{2}\sin\frac{\theta}{2}\sigma_{2}+\frac{1}{2}\sin\frac{\theta}{2}\sigma_{z}\right)$. Red line is the upper bound of the sum of the two variances given by (\ref{usumvar}) and the blue dashed plot denotes the sum of the two variances.}
\label{fig4}
\end{figure}
\begin{equation}\label{usumvar}
\Delta A^2+ \Delta B^2\leq \frac{2\Delta (A-B)^2}{\Big [1-\frac{Cov(A,B)}{\Delta A.\Delta B}\Big ]}-2\Delta A\Delta B.
\end{equation}
As can be seen in Fig. (\ref{fig4}), the bound is actually tight for some classes of qubit states.

{\em Discussions and Conclusions.---}Arguably, the uncertainty relations are the most fundamental relations in quantum theory. It is ironic that after nine decades of the Robertson-Schr{\"o}dinger uncertainty relation, there are still ample scopes to discover tighter uncertainty relations. With the discovery of tighter uncertainty relations, we prove that there is `more' fuzziness in nature than what is allowed by the Heisenberg-Robertson-Schr{\"o}dinger uncertainty relations. 

To summarise, we have derived tighter, state-dependent uncertainty relations both in the sum as well as the product form for the variances of two incompatibles observables. We have also introduced state-dependent reversed uncertainty relations based on variances. Significance of the uncertainty and the reverse relations is that for a fixed amount of `spread' of the distribution of measurement outcomes of one observable, the `spread' for the other observable is bounded from both the sides. These uncertainty relations will play an important role in quantum metrology, quantum speed limits and many other fields of quantum information theory due to the fact that these relations are optimization free, state-dependent and tighter than the most of the existing bounds. On the other hand, reverse uncertainty relations should set the stage for addressing an important issue in quantum metrology, i.e., to set the upper bound of error in measurement and the upper bound for the time of quantum evolutions.

 
\bibliographystyle{h-physrev4}

\end{document}